\documentclass[aps, prl,reprint,groupedaddress,showpacs]{revtex4-1}
\usepackage[version=3]{mhchem}
\usepackage{balance}
\usepackage{graphicx}
\usepackage{lastpage}
\usepackage{mathptmx}
\usepackage{amssymb}
\usepackage{upgreek}
\usepackage{dsfont}
\usepackage{epstopdf}

\begin{document}

\title{Effect of shape and elastic degrees of freedom on the propulsion of bead-spring micro-swimmers}

\author{Jayant Pande}
%\affiliation{
% Institute for Theoretical Physics, Friedrich-Alexander University Erlangen-Nuremberg, Erlangen, Germany
%}
%
%\author{Ulrich R\"ude}
%\affiliation{
%Chair for System Simulation, Friedrich-Alexander University Erlangen-Nuremberg, Erlangen, Germany
%}
%
\author{Ana-Sun\v cana Smith}
\affiliation{
 Institute for Theoretical Physics, Friedrich-Alexander University Erlangen-Nuremberg, Erlangen, Germany
}

%\date{\today}

\begin{abstract}

Bead-based micro-swimmers are promising systems for payload delivery on the micro-scale. However, the principles underlying their optimal design are not yet fully understood. Here we study a simple device consisting of three arbitrarily-shaped beads connected by two springs. We analytically determine the most favorable kinematic parameters for sinusoidal driving, and show how the swimmer changes from being a pusher to a puller. For cargo carrying ellipsoidal beads, we perform geometric optimization under the constraint of a constant total volume or surface area, with the aim of maximizing the device transport velocity and efficiency. Interestingly, we identify two major transport regimes, which arise from the competition between the elastic and the drag forces faced by the swimmer. We construct a phase diagram that indicates when the fastest swimming emerges because of minimized drag, and when due to heightened interactions among the beads.

\end{abstract}

%\pacs{47.63.Gd, 47.63.mh, 87.85.Tu}

\maketitle

In recent years, the motion of self-propelled micro-objects has, in all senses, come under the microscope, with various experimental~\cite{Dreyfus:2005:Nature, Ishiyama:2001:SAA-P, Benkoski:2011:JMaterChem, Li:2012:JApplPhys, Breidenich:2012:SoftM, Baraban:2012:ACSNano, Baraban:2012:SoftM, Keim:2012:PhysFluids, Theurkauff:2012:PRL, Sanchez:2011:Science, Liao:2007:PhysFluids, Leoni:2009:SoftM}, theoretical~\cite{Guenther:2008:EPL, Becker:2003:JFM, Ledesma-Aguilar:2012:EurPhysJE, Avron:2005:NewJPhys, Friedrich:2012:PRL, Fuerthauer:2013:PRL, Leoni:2010:PRL}, and simulation~\cite{Pooley:2007:PRL, Elgeti:2013:PNAS, Zoettl:2012:PRL, Pickl:2012:JoCS, Swan:2011:PhysFluids} studies being performed to investigate their behavior in various environments (for reviews see Refs.~\cite{Lauga:2009:RPP, Golestanian:2011:SoftM}). This is mainly with two complementary objectives in mind: understanding the biomechanics of natural micro-organisms, and designing controllable micro-machines.

It now appears that, contrary to initial belief~\cite{Purcell:1977:AmJPhys}, notions of energy loss and efficiency are relevant at the micro-scale~\cite{Vilfan:2011:PNAS, Vilfan:2012:PRL, Kimura:2009:JExpBiol}, and the optimization of the various facets of a micro-swimmer's motion is important for both the aforementioned objectives. One way to do this is via ``kinematic optimization",  which involves finding the best swimming strokes for a certain swimmer~\cite{Tam:2007:PRL, Pironneau:1974:JFM, Vilfan:2011:PNAS, Avron:2004:PRL, Spagnolie:2010:PhysFluids}. Alternatively, one may pursue the less-traversed path of ``geometric optimization", which concerns finding the best structural parameters for a class of swimmers following the same swimming stroke~\cite{Pironneau:1973:JFM, Vilfan:2012:PRL}. In both approaches, most of the work done so far has been numerical, at least when one considers the final results.

An analytical model which has had a major impact on the understanding of the physics of micro-swimming is the three-sphere swimmer~\cite{Najafi:2004:PRE, Golestanian:2008:PRE}. It consists of three collinear spheres linked by two arms following a defined periodic stroke. Apart from setting the theoretical framework, the utility of this design has been demonstrated in a number of experimental systems based on linearly connected beads~\cite{Dreyfus:2005:Nature, Benkoski:2011:JMaterChem, Li:2012:JApplPhys, Breidenich:2012:SoftM, Baraban:2012:SoftM, Leoni:2009:SoftM}. Today, bead-based assemblies are strongly considered as ideal micro-carriers, since the payload can be naturally placed in the interior of the beads or on their surface.

Given the wide study of this assembly, it is surprising that theoretical analysis has not been adequately extended to the case when the stroke is not predetermined but results from known forces driving the swimmer. In this work, therefore, we develop a model that can account for this situation and extend the design to beads of arbitrary shape. We perform both kinematic and geometric optimization to determine the optimum driving parameters and, for ellipsoidal shapes, the aspect ratios that maximize the payload transport efficiency. This analysis extends our understanding of the interplay between the elasticity and the hydrodynamics involved.
\begin{figure}
\centering
\includegraphics[width=0.3\textwidth]{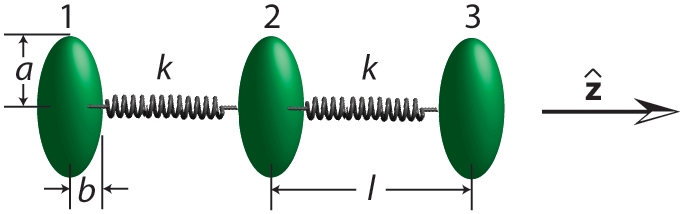}
\caption{(color online) An ellipsoidal swimmer.}\label{fig:ellipsoids}
\end{figure}

Our micro-swimmer (Fig. \ref{fig:ellipsoids}) consists of three identical ellipsoids, formed by revolving an ellipse of semi-axes $a$ and $b$ around $a$, with aspect ratio $e = a/b$. These ellipsoids are arranged with their major axes either parallel or perpendicular to two linear springs of stiffness constant $k$.

To mimic the oscillating nature of the stroke commonly realized in experiments~\cite{Dreyfus:2005:Nature, Li:2012:JApplPhys}, we impose sinusoidal driving forces on the three beads along the swimmer axis ($\mathbf{\hat{z}}$-axis)
\begin{align}\label{eq:dr_forces}
\mathbf F_1^\text{d}(t) &= A \sin\left(\omega t\right) \mathbf{\hat{z}},\nonumber\\
\mathbf F_2^\text{d}(t) &= - \mathbf F_1^\text{d}(t) - \mathbf F_3^\text{d}(t),\text{ and}\nonumber\\
\mathbf F_3^\text{d}(t) &= B \sin\left(\omega t + \alpha\right) \mathbf{\hat{z}},\text{ with }\alpha \in[-\pi, \pi].
\end{align}
Here $A$ and $B$ are non-negative amplitudes of the forces $\mathbf F_1$ and $\mathbf F_3$ applied to the outer beads at the frequency $\omega$ and with the phase difference $\alpha$. The force $\mathbf F_2$ on the middle ellipsoid is defined by the condition for autonomous propulsion, which requires the net driving force on the device to vanish at all times. This force-based protocol is different from that in the original model~\cite{Golestanian:2008:PRE}, where the velocity of a swimmer was calculated with pre-described deformations of the arms. In the approach we employ, the stroke is a consequence of all the forces acting on the bodies, including the ones transmitted by the fluid and the forces originating from the internal degrees of freedom, \emph{i.e.} the spring forces $\mathbf F^\text{s}$.

To determine the velocity of the device, we assume that it moves through an incompressible fluid, the latter governed by the Stokes equation and the incompressibility condition
\begin{equation}
\eta\nabla^2\mathbf u\left(\mathbf r, t\right) - \nabla p\left(\mathbf r, t\right) + \mathbf f\left(\mathbf r, t\right)=0;\,\,\,\,\,\,\nabla\cdot\mathbf u=0.
\end{equation}
Here, $\eta$ is the dynamic viscosity of the fluid at the point $\mathbf r$ at time $t$, moving with a velocity $\mathbf u\left(\mathbf r, t\right)$ under a pressure $p\left(\mathbf r, t\right)$. The force density $\mathbf f\left(\mathbf r, t\right)$ acting on the fluid is given by
\begin{equation}
\mathbf f\left(\mathbf r, t\right)=\sum\limits_{i=1}^3\left[\mathbf F^\text{d}_i(t) + \mathbf F^\text{s}_i(t)\right]\delta\left(\mathbf r - \mathbf R_i(t)\right),
\end{equation}
where the index $i=1, 2, 3$ denotes the $i$-th bead placed at the position $\mathbf R_i$ subject to a net driving force $\mathbf F^\text{d}_i(t)$ and a net spring force $\mathbf F^\text{s}_i(t)$. Assuming no-slip boundary conditions at the fluid-body interfaces, the instantaneous velocity $\mathbf v_i(t)$ of each body~\cite{Doi:1988:OUP} is given by
\begin{equation}\label{eq:v_bodies}
\mathbf v_i=\dfrac{\left(\mathbf F^\text{d}_i + \mathbf F^\text{s}_i\right)}{\gamma} + \sum\limits_{j \neq i}^3\mathbf T\left(\mathbf R_i - \mathbf R_j\right)\cdot\left(\mathbf F^\text{d}_i + \mathbf F^\text{s}_i\right),
\end{equation}
with $\gamma$ being the Stokes drag coefficient~\cite{Perrin:1934:JPhysRadium, Berg:1983:PUP} and $\mathbf T\left(\mathbf r\right)$ the Oseen tensor~\cite{Happel:1965:P-H, Oseen:1927:LeipzigAV}. The latter is here diagonal, due to the collinear nature of the driving forces and the employed far-field approximation ($a/l\ll1$).

Following~\cite{Felderhof:2006:PhysFluids}, the steady state body positions are
\begin{equation}
\mathbf R_i(t)=\mathbf S_{i0} + \boldsymbol \zeta_i(t) + \mathbf vt,
\end{equation}
where $\boldsymbol\zeta_i$ denotes small oscillatory perturbations around the equilibrium configuration $\mathbf S_{i0}$ of the device that moves with a uniform velocity of swimming $\mathbf v$ given by
\begin{equation}\label{eq:v_swim}
\mathbf v=\dfrac{1}{3\tau}\int\limits_0^\tau\sum\limits_{i=1}^3\mathbf v_i(t)\mathrm dt.
\end{equation}
This swimming velocity is obtained by averaging over the time-period $\tau$ required to perform one swimming cycle.

On introducing the dimensionless `reduced spring constant' $\kappa$ and the `reduced hydrodynamic radius' $\lambda$
\begin{equation}\label{eq:reduced}
\kappa=\frac{k}{\pi\eta\omega l} \ \ \ \text{, } \ \ \ \lambda=\frac{1}{l}\frac{\gamma}{6 \pi \eta},
\end{equation}
the perturbation approach allows us to solve eqs. (\ref{eq:v_bodies}) and (\ref{eq:v_swim}) to obtain the swimming velocity to the leading order in $\lambda$,
\begin{equation}\label{eq:v_force}
\mathbf v = \dfrac{7 \lambda\left[A B\left(\kappa^2 + 12\lambda^2\right)\sin\alpha + 2\left(A^2 - B^2\right)\kappa \lambda\right]}{24 l^3 \pi^2 \eta^2\omega\left(\kappa^4 + 40\kappa^2 \lambda^2 + 144 \lambda^4\right)} \mathbf{\hat{z}}.
\end{equation}
This expression can be applied to any device consisting of three identical bodies of a known hydrodynamic radius subject to the force protocol given in eq.~(\ref{eq:dr_forces}). It can be mapped to the formula of Golestanian and Ajdari~\cite{Golestanian:2008:PRE}, $\mathbf v = G d_1 d_2 \omega \sin\beta \mathbf{\hat{z}}$, that relates the velocity to the stroke. Here $G$ is a geometric factor and $d_1=\text{max}\{|\boldsymbol\zeta_2 - \boldsymbol\zeta_1|\}$ and $d_2=\text{max}\{|\boldsymbol\zeta_3 - \boldsymbol\zeta_2|\}$ are the amplitudes of the oscillations of the swimmer's arms, with $\beta$ giving the phase difference between the two. In our formulation these stroke parameters are functions of the swimmer's geometry and the driving parameters.% (see SM for details).

%Interestingly, eq. (\ref{eq:v_force}) shows that the swimmer has a net non-zero velocity even for a time-reversible forcing protocol (\emph{i.e.}, for $\alpha = 0$ or $\pi$), as long as $A \neq B$. This result does not contradict the ``Scallop theorem"~\cite{Purcell:1977:AmJPhys, Ludwig:1930:ZVP}, which prohibits time-reversible motion, because while the forces are reversible in time, the induced arm-length deformations are not.
\begin{figure}
\centering
\includegraphics[width=0.47\textwidth]{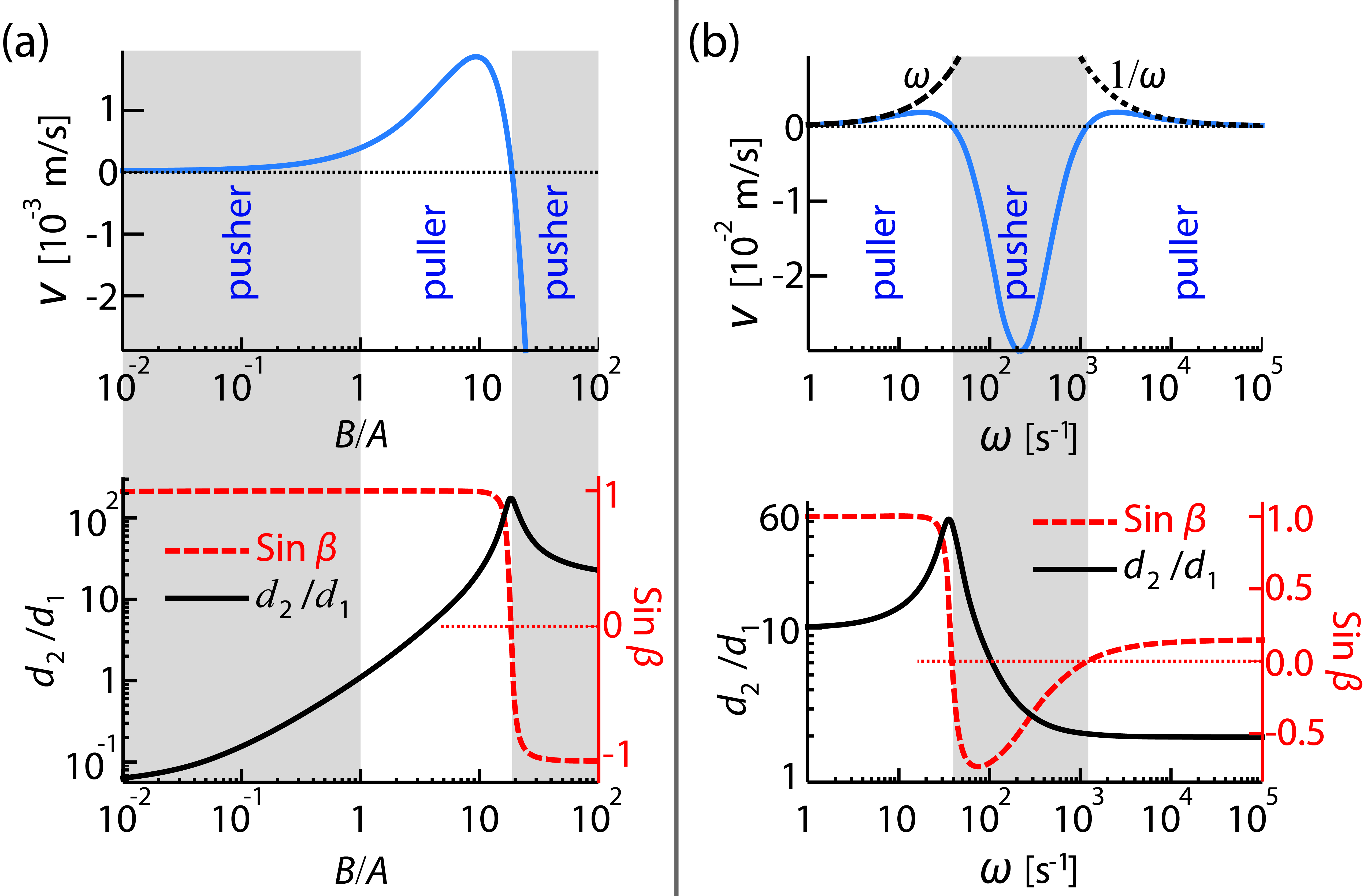}
\caption{(color online) Velocity $\mathbf v$, armlength ratio $d_2/d_1$ and stroke phase difference $\sin\beta$ of a swimmer as a function of (a) force ratio $B/A$, and (b) frequency $\omega$. Here $\alpha = \pi/2$.}\label{fig:kinematic}
\end{figure}

\emph{Kinematic optimization.}---Given a fixed swimmer configuration, we first determine the parameters of driving that lead to the fastest swimming. Clearly from eq. (\ref{eq:v_force}), the optimal phase difference is $\alpha = \pi/2$, if $A \ge B$, while increasing the force amplitudes leads to a quadratic increase in the velocity.

Analysis of eq.~(\ref{eq:v_force}) and of the associated flow-fields shows that one can actually specify if the swimmer is a puller or a pusher by checking the relation
\begin{equation}\label{eq:puller}
\left(\dfrac{B}{A} - \dfrac{A}{B}\right)^{-1}\sin\alpha \gtrless  \dfrac{2 \kappa \lambda}{\kappa^2 + 12 \lambda^2}.
\end{equation}
When the left hand side of relation (\ref{eq:puller}) is larger, then the swimmer moves in the direction of the bead with the higher force amplitude, and the swimmer is consequently a puller. Otherwise it is a pusher. The exception is the limit $\mathbf v \rightarrow \mathbf 0$, when the nature of the swimmer is not clearly defined.

In particular, assuming $\alpha > 0$,  the swimmer is a pusher if $A > B$ (Fig.~\ref{fig:kinematic}a). If $B > A$, yet is small enough for the left hand side of relation (\ref{eq:puller}) to be smaller, then the swimmer is a puller. If $B$ is sufficiently large compared to $A$, the swimmer is again a pusher. In the special case of $\alpha=\pi/2$ this latter transition coincides with the maximum in the $d_2/d_1$ curve (lower panel in Fig.~\ref{fig:kinematic}a). For a changing ratio of $B/A$, due to the quadratic nature of the velocity curve, the global maximum of the velocity is always in the pusher regime. For other parameter changes ($\omega$, $\alpha$, $\kappa$, $\eta$), this does not necessarily hold true.

Interestingly, a pusher can be turned into a puller, and \emph{vice versa}, simply by varying the driving frequency $\omega$ (Fig.~\ref{fig:kinematic}b). In this case, if the other parameters are held fixed, then the sign of the phase shift $\sin\beta$ of the stroke alone determines the pusher/puller nature of the swimmer. We find that $\mathbf v \sim \omega$ for $\omega \rightarrow 0$ and $\mathbf v \sim 1/\omega$ for $\omega \rightarrow \infty$. This is in contrast to the linear dependence of the swimming velocity $\mathbf v$ on the driving frequency $\omega$ for all $\omega$, when the stroke is pre-set~\cite{Golestanian:2008:PRE}.

Another intriguing effect is the near locking of the stroke phase shift $\beta$ for large parts of the parameter space (lower panel of Fig.~\ref{fig:kinematic}a). This suggests that making the force amplitudes flexible allows the swimmer to automatically synchronize its two beating arms so as to achieve efficient propulsion. This is reminiscent of the phase locking observed in \emph{Chlamydomonas} flagella when elastic connections are included~\cite{Leptos:2013:PRL}.

\begin{figure}
\includegraphics[width=0.9\columnwidth]{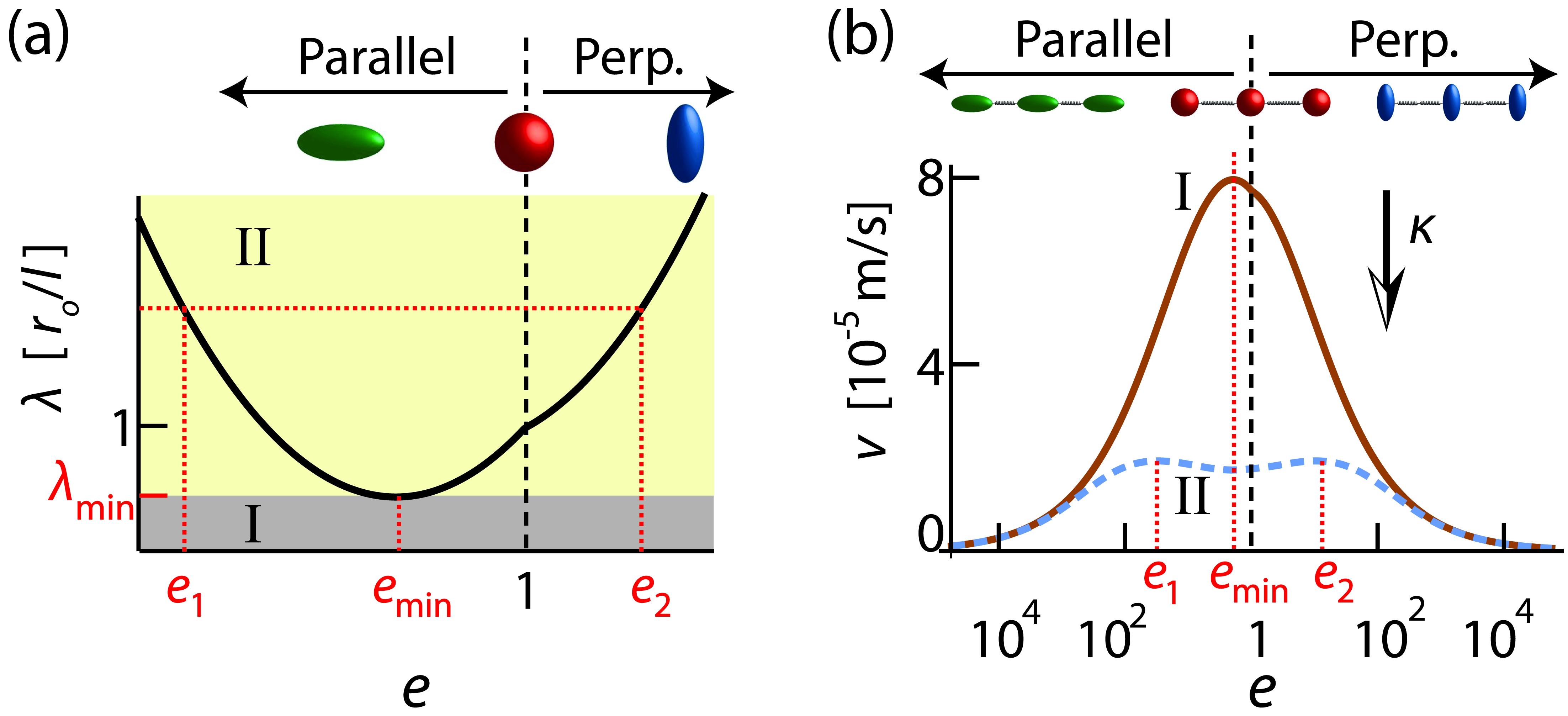} \label{fig:Fig3_b}
\caption{(color online) For prolate ellipsoids of a constant volume, (a) reduced radius $\lambda$, and (b) typical velocity $\mathbf v$ curves, for $A = B$ or $\alpha = \pm \pi/2$ \& $A \gtrless B$. The drag dominated regime (`I') and the interaction-dominated regime (`II') are indicated.}
\label{fig:Fig3}
\end{figure}

\emph{Geometric optimization.}---Given a fixed driving protocol, we now determine the precise shapes of the ellipsoidal beads that lead to the fastest cargo transport. Assuming that a fixed quantity of payload is carried either within the beads or on their surface, we impose a constraint on the bead volume or area. We keep the aspect ratio $e$ of the individual ellipsoids ($e = a/b$) as a free parameter ($e = 1$ denotes a sphere of radius $r_0$), and look at the equation $\mathrm d\left|\mathbf v\right|/\mathrm de = 0$. We find it useful to separate this into two components using the chain rule, $\mathrm d\lambda/\mathrm de = 0$ and $\mathrm d\left|\mathbf v\right|/\mathrm d\lambda = 0$.

Solving $\mathrm d\lambda/\mathrm de=0$ for $e$ yields the aspect ratio $e_\text{min}$ of the ellipsoid with the smallest effective hydrodynamic radius $\lambda_\text{min}$, \emph{i.e.} the most streamlined shape, under the relevant constraint (Table~\ref{tab:eto}). Since this condition relates only to the geometry of the beads and not to the forces acting on the swimmer, the velocity curve always has an extremum $\mathbf v_{\lambda_\text{min}}$ at the aspect ratio $e_\text{min}$ (Fig.~\ref{fig:Fig3}b).
 Depending on the reduced spring constant $\kappa$, this extremum may be a maximum or a minimum.

The equation $\mathrm d\left|\mathbf v\right|/\mathrm d\lambda=0$ allows us to connect the optimal shapes to the different forces acting on the beads, since its solutions relate the effective radius to the spring constant, the driving frequency and the fluid viscosity. More specifically, the equation has a solution $\lambda_1 = \kappa/(2\sqrt 3)$, which is always real and positive. This leads to the velocity
\begin{equation}\label{eq:v_max1}
\mathbf v_{\lambda_1}=\dfrac{7\left(2\sqrt 3 A B \sin\alpha + A^2 - B^2\right)}{768\, \pi^2\eta^2 l^3 \omega \kappa} \mathbf{\hat{z}}.
\end{equation}

\begin{table}\caption{Critical values of $e_\text{min}$ and their respective $\lambda_\text{min}$ (scaled by $r_0/l$), for prolate and oblate ellipsoids subject to a constant total volume ($V$) or surface area ($S$) constraint.}\label{tab:eto}
\begin{tabular}{  c | c c | c c }

             & $V=\text{const}$ &           & $S=\text{const}$ & \\ \hline
             & $e_\text{min}$ & $\lambda_\text{min}$  & $e_\text{min}$   & $\lambda_\text{min}$ \\ \hline
    Prolates & $1.95$    & $0.95$  & $4.04$    & $0.89$ \\ \hline
    Oblates  & $0.70$    & $0.99$  & $0.00$    & $0.80$ \\

  \end{tabular}
\end{table}

\begin{figure}
\centering
\includegraphics[width=0.45\textwidth]{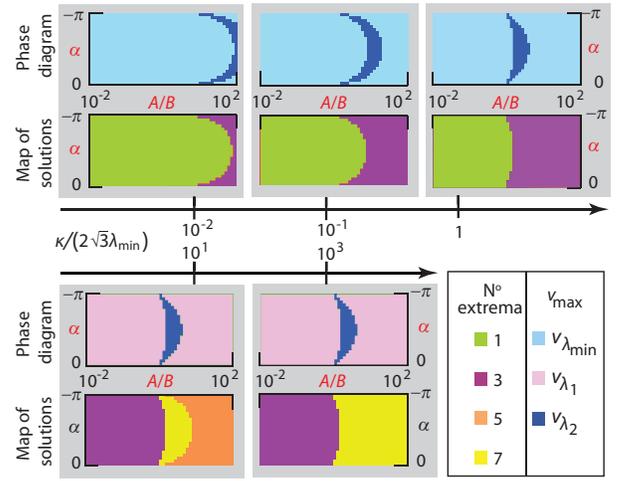}
\caption{(color online) $A/B$ vs $\alpha$ phase diagrams for different values of $\kappa$. For $\kappa/\lambda_\text{min} < 2 \sqrt 3$, for some parameters we have $\mathbf v_\text{max} = \mathbf v_{\lambda_\text{min}}$ (\emph{i.e.} the drag-dominated regime), and for $\kappa/\lambda_\text{min} > 2 \sqrt 3$ the whole phase space is interaction-dominated.% Note that the system is invariant under the transformation $\{A \rightarrow B,\,B \rightarrow A,\,\alpha \rightarrow -\alpha,\,\mathbf{\hat{z}} \rightarrow -\mathbf{\hat{z}}\}$.
}\label{fig:phase_diag}
\end{figure}

For the special cases of $A = B$ or $\alpha = \pm \pi/2$ (for $A \gtrless B$), $\lambda_1$ is the only solution. If the springs are so soft that $\kappa<2\sqrt 3 \lambda_\text{min}$, then there can be no ellipsoid with the effective radius $\lambda_1$, since $\lambda_1 < \lambda_\text{min}$ (regime I in Fig.~\ref{fig:Fig3}a). The velocity curve then has only one extremum obtained from the geometric condition (regime I in Fig.~\ref{fig:Fig3}b). For oblates of a constant volume and prolates, on the other hand, if the springs are stiff enough so that $\kappa > 2\sqrt 3 \lambda_\text{min}$, then exactly two ellipsoids have the effective radius $\lambda_1$ (with aspect ratios given by $e_1$ and $e_2$ in regime II, Fig.~\ref{fig:Fig3}a). Consequently, in addition to the extremum obtained at $e = e_\text{min}$, the velocity curve has two more extrema, leading to degenerate velocity values ($\mathbf v|_{e_1} = \mathbf v|_{e_2} = \mathbf v_{\lambda_1}$, regime II, Fig.~\ref{fig:Fig3}b), with $\left|\mathbf v_{\lambda_1}\right| > \left|\mathbf v_{\lambda_\text{min}}\right|$. The case of oblates of a constant surface area (discussed in the Supplemental Material) is different in the details since $e_\text{min}=0$.

The two regimes in Fig.~\ref{fig:Fig3} are distinguished not just by the number of extrema, but also by the nature of the largest absolute velocity, which is a manifestation of the interaction between the swimmer's elastic degrees of freedom and the drag force. Specifically, in the so-called `drag-dominated' regime I, the maximum velocity $\left|\mathbf v_\text{max}\right|$ is $\left|\mathbf v_{\lambda_\text{min}}\right|$, achieved by the most streamlined shape (\emph{i.e.} at $e = e_\text{min}$). In contrast, the `interaction-dominated' regime II seems to invert the naive expectation of the drag force hindering motion through the fluid as $\left|\mathbf v_{\lambda_\text{min}}\right|$ is locally the smallest velocity, and $\left|\mathbf v_\text{max}\right| = \left|\mathbf v_{\lambda_1}\right|$. 
% is attained by ellipsoids whose effective radius ($\lambda_1$) relates the system viscosity, elastic stiffness and the hydrodynamic drag.
 These two regimes emerge because the drag has two conflicting effects upon a swimmer: while it resists motion through the fluid, it also promotes the fluid's agitation, resulting in hydrodynamic interaction among the beads and ultimately in swimming. In the interaction-dominated regime, where the spring constant (and consequently $\kappa$) is relatively high, most of the input work is consumed in deforming the springs, and so an increased drag is beneficial for a heightened hydrodynamic interaction among the bodies. Therefore, the swimmer with ellipsoids of an effective radius $\lambda_\text{min}$, which agitates the fluid the least, is locally the slowest. In contrast, in the drag-dominated regime, where the spring constant is low, most of the input work is transferred directly onto the agitation of the fluid, so having a high drag only slows the swimmer down.

\emph{Phase diagram.}---For a general choice of parameters, $\mathrm d\left|\mathbf v\right|/\mathrm d\lambda=0$ provides two further pairs of solutions, namely $\lambda_{2,3}$ (for $B < A$, with $\lambda_2 < \lambda_1 < \lambda_3$) and $\lambda_{4,5}$ (for $A < B$, with $\lambda_4 < \lambda_1 < \lambda_5$). Each physically relevant $\lambda_i$ ($\lambda_i \in \mathds R$ and $\lambda_i \ge \lambda_\text{min}$) provides two degenerate velocity extrema $\mathbf v_{\lambda_i}$. Furthermore, the degeneracy extends over the solutions pairs, with $\mathbf v_{\lambda_2}=\mathbf v_{\lambda_3}$ and $\mathbf v_{\lambda_4} = \mathbf v_{\lambda_5}$. These extrema are given by
\begin{equation}\label{eq:v_max2}
|\mathbf v_{\lambda_i}| = \dfrac{7\left(F_>^2 - F_<^2 - \sqrt{F_>^4 + F_<^4 - 2 F_>^2 F_<^2 \cos\left(2\alpha\right)}\right)}{384\, \pi^2 \eta^2 l^3 \omega \kappa},
\end{equation}
where $i = 2, ..., 5$ and $F_>$ and $F_<$ denote the larger and the smaller of $A$ and $B$, respectively. Consequently, $\mathbf v$ as a function of $e$ has, in addition to one extremum from $\lambda_\text{min}$, up to 3 pairs of extrema from $\lambda_i$. We construct phase diagrams showing the number of velocity extrema (bottom graph in each panel in Fig.~\ref{fig:phase_diag}) and the extremum with the largest absolute value (top graph) as a function of the driving parameters, and for increasing values of $\kappa$. Since the velocity magnitude is unchanged under the transformation $\{A \rightarrow B,\,B \rightarrow A,\,\alpha \rightarrow -\alpha\}$, we restrict the phase diagram to $-\pi\le\alpha\le0$. For $A/B<1$ only $\lambda_1$ and $\lambda_\text{min}$ may be valid, while for $A/B>1$, the relevant solutions may include $\lambda_1$, $\lambda_2$, $\lambda_3$ and $\lambda_\text{min}$. Symmetrically, for $0\le\alpha\le\pi$, $\lambda_4$ and $\lambda_5$ replace $\lambda_2$ and $\lambda_3$.
%$\lambda_{4,5}$ replace $\lambda_{2,3}$.

We investigate the evolution of the phase diagram as the swimmer's springs increase in stiffness from very soft (upper left panel in Fig.~\ref{fig:phase_diag}) to very hard (bottom right panel). For $\kappa = 0$, the swimmer is always in the drag dominated regime (light blue regions, top graphs) irrespective of the other parameters, and there is only one extremum in the velocity curve (indicated by light green regions, bottom  graphs), given by $\left|\mathbf v_\text{max}\right| = \left|\mathbf v_{\lambda_\text{min}}\right|$. At small $\kappa$ (left and middle gray panels in the top row), two other maxima, associated with $\lambda_2$, may appear in the velocity curve (purple regions, bottom graphs). For driving parameters ($A/B$ and $\alpha$) associated with the crescent-shaped dark blue region in the phase diagram, $\left|\mathbf v_\text{max}\right| = \left|\mathbf v_{\lambda_2}\right|$, and the swimmer is in the interaction-dominated regime. This blue region progressively moves towards smaller $A/B$ ratios for increasing values of $\kappa$, till $\kappa = 2\sqrt 3\lambda_\text{min}$. At this critical $\kappa$ value, two more extrema, associated with $\lambda = \lambda_1$, appear in the velocity curve across the entire parameter range. This coincides with the abrupt disappearance of the drag-dominated regime from the phase diagrams, as $\left|\mathbf v_\text{max}\right|$ is either $\left|\mathbf v_{\lambda_1}\right|$ (pink regions, top graphs) or $\left|\mathbf v_{\lambda_2}\right|$. This complete dominance of the interaction-dominated regime for $\kappa > 2\sqrt 3\lambda_\text{min}$ is echoed in the earlier-seen transition between the two regimes (Fig.~\ref{fig:Fig3}b). For higher $\kappa$, the velocity curve gets more extrema (yellow regions in the map of solutions), due to $\lambda = \lambda_3$, but the nature of $\left|\mathbf v_\text{max}\right|$ remains unchanged. The overall shape of the phase diagram becomes independent of $\kappa$ (bottom row), and the swimmer is always in the interaction-dominated regime.

\begin{figure}
\centering
\includegraphics[width=0.4\textwidth]{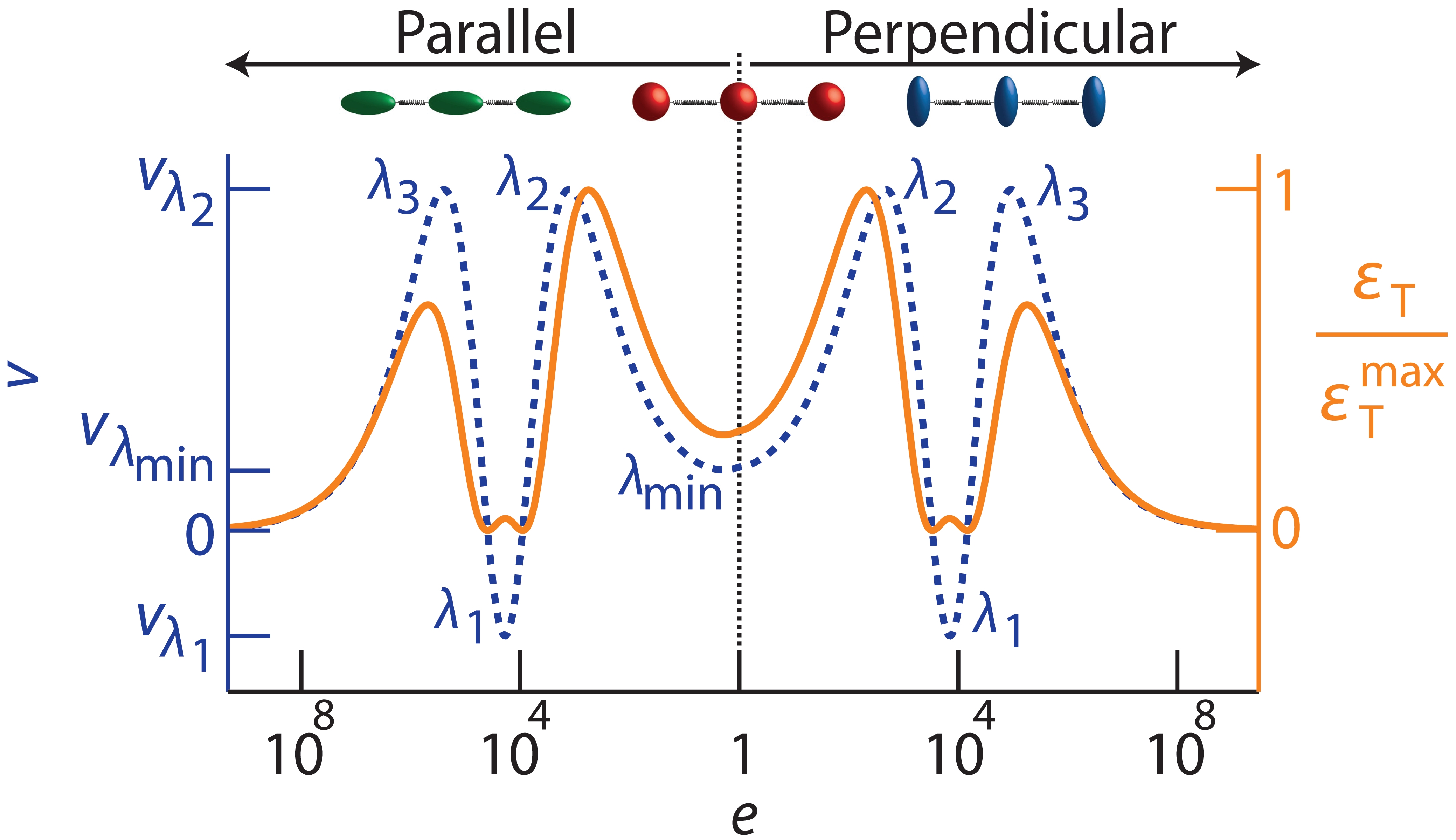}
\caption{(color online) Velocity $\mathbf v$ and transport efficiency $\epsilon_\text{T}$ of constant-volume prolate swimmers, with $B = 3 A$ and $\alpha = \pi/4$.}\label{fig:effT}
\end{figure}
\emph{Transport efficiency.}---To quantify the ability of swimmers to carry cargo, we define the transport efficiency $\epsilon_\text{T}$ as the ratio of $\left|\mathbf v\right|^2$ and the input power $\frac{1}{T}\int_0^T \sum_{j=1}^3 \mathbf F_j(t)\cdot\mathbf v_j(t)\mathrm dt$, giving
\begin{align}
\epsilon_\text{T} = \left|\mathbf v \dfrac{A B  \left(\kappa^2 + 12\lambda^2\right)\sin\alpha - 2\left(B^2 - A^2\right)\kappa\lambda}{\left(A^2 + B^2\right)\left(\kappa^2 + 12\lambda^2\right) - A B \left(\kappa^2 - 12\lambda^2\right)\cos\alpha}\right|.\nonumber
\end{align}
This definition favours fast swimmers, but penalizes ones which require a high power input. It is also bounded as a function of $\omega$, $\kappa$ and $\lambda$, thus ensuring that it does not diverge on, for example, increasing the time period. It is more suitable than the simple ratio of the current ($\propto \mathbf v$) to the input work (as in~\cite{Felderhof:2006:PhysFluids}), which is insensitive to changes in shape for fixed driving. Also, the Lighthill efficiency~\cite{Lighthill:1952:CPAM} is unsuitable because it penalizes swimmers which face a high drag, which is inapt for the interaction-dominated swimming regime.% Using our definition, we find that the fastest swimmers generally have relatively high efficiencies, although locally the most efficient swimmer is not necessarily the fastest one (Fig.~\ref{fig:effT}).%, as faster swimmers also consume more input energy. 

In spite of the natural correlation between the transport velocity and efficiency, the most efficient swimmer is not necessarily the fastest one (Fig.~\ref{fig:effT}). This is particularly important in the interaction dominated regime, where designs which propagate with the same speed can have significantly different efficiencies due to a different repartition of the input work on the fluid and the compression of springs.  For instance, in Fig.~\ref{fig:effT}, $\epsilon_\text{T}$ at $\lambda_3$ is much less than at $\lambda_2$, although $\mathbf v_{\lambda_2} = \mathbf v_{\lambda_3}$. In contrast, in the drag-dominated regime, the input work consumed by the elastic components is negligible, and so optimally shaped swimmers are typically the most efficient.

\emph{Conclusion.}--- Here we studied the effect of the interplay between elastic and hydrodynamic forces on micro-swimming, in the case of a linear swimmer composed of beads of arbitrary shape connected by elastic springs. We showed that starting from an \emph{a priori} fixed driving protocol leads to several important effects including phase locking and conversion from pullers to pushers, and allows the determination of the optimal swimmer shapes. Most importantly, we identified two regimes of transport, one where the fastest swimming occurs when the drag on the bodies is minimal, and the other when the swimming is promoted by strong interactions between the bodies. While the simple geometry allowed us to analytically quantify these effects for ellipsoidal swimmers transporting payload, these effects should be general, and occur whenever the swimmer has elastic degrees of freedom.

\emph{Acknowledgment.}---We gratefully acknowledge support by the Cluster of Excellence: Engineering of Advanced Materials at the Friedrich-Alexander University in Erlangen, Germany. We also thank U. R\" ude, K. Pickl, H. K\"ostler and K. Mecke for valuable comments and discussions.

\bibliography{rsc3} %your .bib file
\bibliographystyle{rsc3}

\end{document}